\documentclass{article}

\usepackage{multirow}
\usepackage{PRIMEarxiv}
\usepackage{amsmath}
\usepackage[utf8]{inputenc} 
\usepackage[T1]{fontenc}    
\usepackage{hyperref}       
\usepackage{url}            
\usepackage{booktabs}       
\usepackage{amsfonts}       
\usepackage{nicefrac}       
\usepackage{microtype}      
\usepackage{lipsum}
\usepackage{fancyhdr}       
\usepackage{graphicx}       
\graphicspath{{media/}}     

\title{Analyzing long-term rhythm variations in Mising and Assamese using frequency domain correlates
\thanks{\textit{\underline{Citation}}: 
\textbf{Authors. Title. Pages.... DOI:000000/11111.}} 
}

\author{
  Parismita Gogoi \\
  IIT Guwahati, India \\
  DUIET, Dibrugarh University, India \\
  \texttt{parismitagogoi@iitg.ac.in} \\
 \And
   Priyankoo Sarmah \\
  IIT Guwahati, India \\
  \texttt{priyankoo@iitg.ac.in} \\
 \And
     S.R Mahadeva Prasanna \\
  IIT Dharwad, India \\
  IIIT Dharwad, India \\
  \texttt{prasanna@iitdh.ac.in} \\
}

\begin{document}
\maketitle

\begin{abstract}
The current work explores long-term speech rhythm variations to classify Mising and Assamese, two low-resourced languages from Assam, Northeast India.  We study the temporal information of speech rhythm embedded in low-frequency (LF) spectrograms derived from amplitude (AM) and frequency modulation (FM) envelopes. This quantitative frequency domain analysis of rhythm is supported by the idea of rhythm formant analysis (RFA), originally proposed by Gibbon~\cite{Gibbon2019quantify}.  We attempt to make the investigation by extracting features derived from trajectories of first six rhythm formants along with two-dimensional discrete cosine transform-based characterizations of the AM and FM LF spectrograms. The derived features are fed as input to a machine learning tool to contrast rhythms of Assamese and Mising. In this way, an improved methodology for empirically investigating rhythm variation structure without prior annotation of the larger unit of the speech signal is illustrated for two low-resourced languages of Northeast India.
\end{abstract}

\keywords{rhythm; rhythm formant; Mising; Assamese; AM; FM.}

\section{Introduction}
The rhythm formant analysis (RFA) introduces the rhythm spectrogram as a tool to characterize long-term speech rhythm. This technique leverages the temporal information embedded within the amplitude modulation (AM) and frequency modulation (FM) envelopes of speech utterance~\cite{gibbon2021jipa}. In our study, we investigate the variations of speech rhythm of Mising and Assamese, two low-resourced languages from Assam in North-East India, using the RFA methodology. This method does not need manual annotation of speech units, a process that is typically time-consuming and requires specific linguistic expertise. This aspect is crucial, especially when dealing with spontaneous speech. Moreover, the automatic segmentation of speech units using speech technology may not be available for low-resource languages. Our study aims to apply the RFA technique to spontaneous speech samples from Mising and Assamese, using a low-frequency (LF) spectrogram to extract rhythmic cues to differentiate between Mising and Assamese. The RFA was introduced as a pioneering and accessible method, inviting further exploration by the speech research community.

From the time domain perspective of rhythm assessment techniques, speech rhythm metrics provide a means to quantify the temporal characteristics of rhythm inherent in both speech production and perception. The domain of speech rhythm in languages can be broadly divided into stress-timed, mora-timed, and syllable-timed~\cite{Nespor2019} \cite{RAMUS} \cite{moraJ} \cite{murty2007perceptual} \cite{dihingia2020rhythm}. English and German fall under the stress-timed category, characterized by their diverse syllabic constructs, often leading to the reduction of unstressed syllables. On the other hand, Spanish and French exemplify syllable-timed rhythms, marked by their consistent and straightforward syllabic patterns with minimal reduction of unstressed syllables~\cite{dauer1983stress}. Rhythm metrics provide measures to classify languages based on their vocalic and consonantal patterns. These measures take into account the duration of vowels and consonants, rather than focusing solely on syllables or stress. As mentioned, Spanish, a syllable-timed language, exhibits simpler consonant patterns and less vowel variation than English, a stress-timed language~\cite{dauer1983stress}. Ramus and his colleagues introduced methods to quantify these vocalic and consonantal variations~\cite{RAMUS}. Grabe and Low expanded on this by introducing pairwise variability indices, nPVI and rPVI~\cite{GrabeLow} \cite{low}. Other important metrics include coefficients of variation for consonantal intervals (VarcoC)\cite{dellwo2003} and for vocalic intervals (VarcoV)\cite{ferragne2004rhythm}. However, it is found that there is significant variation in rhythm metrics among different speakers~\cite{Dellwo2015jasa}.

From a linguistic perspective, while rhythm metrics are valuable, they do come with challenges. They rely heavily on manual annotation. They are often criticized for being subjective, and prone to human error, especially with the manual measurement of long speech recordings. Traditionally, Praat is being used to annotate phonetic units, a process which can be both time-intensive and complex~\cite{praat}. Although there have been attempts at automated segmentation~\cite{wiget}, they have had different challenges, particularly since forced alignment is not available for all languages~\cite{Loukina}. Moreover, biases in segmentation can arise based on language expertise, further complicating the process~\cite{Loukina}. 

As rhythm metrics, traditionally rooted in the time domain, have been faced with various challenges, it has prompted a shift in focus to frequency domain bottom-up approaches~\cite{Todd1994} \cite{cummins1999language} \cite{traunmuller1994conventional}. In these techniques, speech amplitude envelopes are extracted to analyze the spectral properties of rhythm in the low-frequency (LF) spectrum. Contrasting opinions on time-domain and frequency-domain approaches can be found in the literature~\cite{Kohler2009} \cite{Arvanitiusefulness2012}. Many scholars argue that to solve the multifaceted nature of rhythm, we need to move away from singular dimensional analysis of speech waveforms as it is a time series phenomenon~\cite{Kohler2009} \cite{CUMMINS1998145}. It is hypothesized that characteristics of rhythm lie in the periodicity of the envelope~\cite{Tilsen2013}. Inductive methods offer a streamlined process compared to earlier annotation-based techniques. This is rooted in the belief that natural speech rhythm links to neural resonance patterns, specifically in low frequencies below 10 Hz~\cite{nature} \cite{ding2017temporal}. The interplay between rhythm theories, speech production, and perception models emphasizes the influence of vocal tract filter functions on oscillating amplitude modulations~\cite{CUMMINS1998145} \cite{barbosa2002explaining} \cite{inden2012rapid} \cite{O'Dell}.

Recent studies have highlighted the effectiveness of Gibbon's rhythm formant analysis (RFA)~\cite{Gibbon2019quantify} \cite{gibbon2021jipa}, which basically computes rhythm information from the LF spectrum analysis of both AM and FM envelopes. In this approach, rhythm is viewed as embedded in the modulation output of a carrier signal originating in the larynx region of the vocal tract. The AM envelope is extracted to understand the changes occurring due to higher amplitude vowels and lower amplitude consonants~\cite{Todd1994} \cite{Tilsen2008LowfrequencyFA}.  AM is related to the syllable sonority outline of the waveform~\cite{gibbonComp}, and FM represents variations in linguistic features such as lexical tone, accent, and intonation. The RFA approach particularly concentrates on the spectral range below 10 Hz and is termed as the LF spectral region. RFA illustrates rhythm in terms of frequency values rather than the duration of speech units. It focuses on two primary aspects of physical properties in rhythm, spectral frequency, and spectral amplitude. It defines the concept of rhythm formants (R-formants), which are the dominant magnitude frequencies in the LF spectrum of AM and FM speech envelopes. The RFA method does not require text annotation of the speech, making it possible to use for low-resource languages without the need of text resources. 

R-Formants, when analyzed, reveal distinct rhythmic differences between languages, as evidenced in contrasts between English and Mandarin rhythms~\cite{gibbon2018future}. In-depth examinations of rhythm formants and rhythm spectrograms from amplitude envelopes are performed in various languages~\cite{gibbon2020storyreading}. Subsequent analyses highlight the distinct rhythmic patterns in extended readings across diverse languages, including a comprehensive classification of the Edinburgh fable database~\cite{gibbon2021jipa}. As a recent development, the concept of rhythm spectrogram has been introduced in RFA to characterize long-term rhythm using the temporal information inherent in the AM and FM envelopes of speech utterances~\cite{gibbon2021jipa}. The modulation-theoretic rhythm spectrogram works as an inductive method to capture the time-varying nature of rhythm. Authors~\cite{kaustubh2023rhythm} explore the RFA in the depression speech, and show how features derived from R-formants can be used to classify depression speech from its non-depression counterpart.   

The current study focuses on analyzing long-term rhythmic patterns in spontaneous speech utterances when individuals describe predetermined subjects in their respective native languages. The rhythm formant and the rhythm spectrogram derived from the AM and FM envelopes and the highest magnitude frequency vector through the spectrogram are explored. Temporal information about long-term rhythm is investigated using the trajectory of the first six highest-amplitude dominant frequencies.  

\subsection{Motivation and contributions}
Rhythm variations of Assamese and Mising are compared in the present study using LF rhythm spectrograms. Mising language belongs to the Tibeto-Burman language family with eight dialects~\cite{taid}. Oyan, Pagro, and Delu dialect speakers reside mainly in Assam, and Sayang, Dambuk, Somua, Moying, and Tamar dialect speakers reside in the hills and foothills of Arunachal. Samuguria is not based on any dialect as these speakers living in Assam have abandoned speaking Mising and have adopted Assamese~\cite{Prasad}. The Pagro dialect of Mising is spoken in Upper Majuli and Dhemaji, while the Delu dialect is spoken on the southern bank of Brahmaputra. Assamese, on the other hand, is the official language of Assam, spoken by about 15 million Assamese people and by many Tibeto-Burman speakers in Assam as a lingua-franca. Assamese is an Indo-Aryan language with three main recognized dialects, namely, Eastern Assamese (EA), Central Assamese (CA), and Western Assamese (WA)~\cite{Goswami-and-Tamuli-2003}. The Mising tribe has been residing near the Assamese in the Brahmaputra valley of Assam, resulting in contact-induced linguistic changes in both Mising and Assamese. As Assamese is the official language of Assam and is used extensively in education and administration, the Misings have become `semi-fluent' bilinguals in Mising and Assamese~\cite{Rajeev}. As a result, Assamese has affected the lexicon, grammar, and sound system of Mising to a considerable extent~\cite{Loreina}. This influence may not be uniform; it varies depending on the extent of exposure to Assamese and the specific geographic locations of the Mising subgroups. An exciting avenue of research is the exploration of these rhythmic variabilities in the context of rhythm formants. In this study, we aim to achieve two primary objectives. First, we seek to characterize and classify the speech rhythms of both Assamese and Mising. Secondly, we intend to analyze the Pagro and Delu dialects to determine which is more closely aligned with Assamese in terms of rhythm.

To the best of our knowledge, apart from a temporal-based~\cite{parismita_specom} and a RFA-based~\cite{parismitatallip} rhythm study on Mising done by us, there has been no other research in this area. In our previous work, a cross-linguistic rhythm measures analysis of Mising and Assamese was conducted~\cite{parismita_specom}. The experiment involved the automatic detection of vocalic and non-vocalic regions leading to the computation of conventional rhythm measures such as \%V, n-PVI, r-PVI, $\Delta$V etc. The results of the study showed that the conventional rhythm measures can classify between Assamese and the two Mising dialects with the best accuracy of 83.10\%. While such results are encouraging, automatic detection of segmental boundaries for measuring rhythm may not always be accurate. Moreover, in our previous work on analyzing R-formants for Mising and Assamese~\cite{parismitatallip}, we explored the LF spectrum of AM envelopes to develop cues for Mising and Assamese rhythm discrimination. The LF spectrum provides a gross view of the rhythm of the entire speech utterance. However, it does not indicate how the rhythm might vary over time~\cite{gibbon2021jipa}. The LF spectrogram is studied in the present work to discriminate Mising and Assamese speech rhythm variations. This method does not require manual annotation of speech units~\cite{gibbonRhythmZoneTheory}, which is important for spontaneous speech and low-resource languages where automatic segmentation may not be available. Moreover, the analysis of rhythm variation could prove to be highly beneficial for long-duration speech files, particularly those related to storytelling, long conversations, etc. The LF spectrogram approach has been used to investigate rhythm patterns in extended readings across various languages~\cite{gibbon2021jipa}.

The LF rhythm spectrogram is a two-dimensional representation of LF spectrum windowed speech stacked temporally, and the prosodic structure is derived from the LF spectrogram of the speech signal. In the current work, temporal information about long-term rhythm properties is extracted, and support vector machine (SVM) based classifiers are designed to explore the effects of Assamese rhythm over two dialects of Mising, namely Pagro and Delu. Long-term rhythm dynamics are investigated for longer-duration utterances in which speakers narrate given topics in their native languages. The AM and FM-based envelopes are used to derive the LF spectrograms, from which R-formant trajectories are computed. The variance of six dominant R-formants is explored as cues characterize the long-term rhythm variations. In addition to that, two-dimensional discrete cosine transform (2D-DCT) coefficients of the LF spectrogram are extracted to utilize as potential cues for discriminating spontaneous speech narratives in Assamese vs. Pagro and Assamese vs. Delu using SVM classifiers.

The current paper is organized as follows. Section~\ref{database} presents a detailed description of database preparation, and details of the experiments performed are mentioned in Section~\ref{experiments}. Section~\ref{results} discusses the results of statistical analysis and SVM-based classification. Section~\ref{discussion} concludes the paper by summarizing the work and providing future directions.

\section{Database preparation}\label{database}
In this section, we discuss the database preparation process and experimental details in two subsections. 
\begin{table}[t]
\centering
\caption{Dataset details.}
\label{tab:database}
\resizebox{\textwidth}{!}{
\begin{tabular}
{c|cc|c|c|cc}
\hline
{Group} & \multicolumn{2}{c|}{Gender}                           & {Age ($\mu$ $\pm$ $\sigma$)} & {number of utterances} & \multicolumn{2}{c}{Duration}                 \\ \cline{2-3} \cline{6-7} 
                                               & \multicolumn{1}{c|}{M} & F                            &                                                          &                                            & \multicolumn{1}{c|}{Total (min)} & Average (sec)          \\ \hline
Assamese (A)         & \multicolumn{1}{c|}{1} & 4                            & 22.55 $\pm$ 1.25                                         & 108                                        & \multicolumn{1}{c|}{24.78}      & 13.76 $\pm$ 2.00 \\ \hline
Mising - Pagro (P)   & \multicolumn{1}{c|}{2} & 5                            & 34.31 $\pm$ 7.69                                         & 142                                        & \multicolumn{1}{c|}{30.41}      & 12.85 $\pm$ 2.12 \\ \hline
Mising - Delu (D)    & \multicolumn{1}{c|}{2} & 5                            & 30.07 $\pm$ 11.14                                        & 158                                        & \multicolumn{1}{c|}{32.84}      & 12.47 $\pm$ 1.89 \\ \hline
Total                & \multicolumn{1}{c|}{5} & 14 & 19 (Total speaker)             & 408             & \multicolumn{1}{c|}{88.03}     & 12.94 $\pm$ 2.06 \\ \hline
\end{tabular}}
\end{table}

\subsection{Speakers}
A speech dataset was collected from 19 speakers of the Assamese and Mising languages. This dataset aimed to study long-term rhythm variations in spontaneous speech narratives. The speakers recorded were from the Jorhat variant of the Eastern Assamese (A) dialect in the upper Assam region. This dialect is considered the standard Assamese as mentioned by Goswami and Tamuli~\cite{Goswami-and-Tamuli-2003}. Seven Pagro Mising speakers, who mostly live near Simen Sapori in the Dhemaji district of Assam, were also recorded. There are Delu Mising speakers who live near the Burhi-dihing river in the Rajabari area of the Sivasagar district. The distance between Simen Sapori and Rajabari is about 150 KM. All recordings were made using a Zoom H1n recorder in a quiet environment. The audio was recorded at a sampling frequency of 44.1 kHz with a 16-bit depth in .WAV format. The data collection took place from January 2020 to January 2021. All 19 speakers confirmed they had no speech or language disorders during the recording. Further details about the speakers and the database are provided in Table~\ref{tab:database}. Most participants in this study were bilingual or trilingual. Mising participants spoke combinations of Assamese, Hindi, and English. Assamese participants spoke either Hindi or English but not Mising.

\subsection{Materials}
In this study, we primarily used spontaneous speech narratives for elicitation. We provided five topics to speakers from both languages. Not every speaker discussed all five topics; they chose based on their comfort and available time. Typically, a speaker covers 2-3 of these topics. These topics revolved around the daily lives and cultural practices of the Assamese and Mising communities. For Assamese speakers, the topics were: the Bihu festival, an introduction to their village, the process of weaving Assamese garments, the Assamese community, and traditional food preparation methods. For Mising speakers, the subjects were: the Ali-Aye-Ligang festival, an introduction to their village, the process of weaving Mising garments, the Mising community, and traditional food-making techniques. Recorded speeches varied in duration, ranging from 1 to 4 minutes. We reviewed these recordings, identified and marked sentence boundaries by hand, discarding any portions with noise. Table~\ref{database} displays the total and average speech duration for each language. For the experiments in subsequent sections, we utilized sentence-level audio clips for analysis and modeling.

\section{Experimental details}\label{experiments}
This section discusses the experimental details of the current works. The procedure of extracting the AM and FM envelope-based LF spectrograms is discussed first, and then we discuss the feature computation process using two-dimensional discrete cosine transform (2D-DCT). The linear mixed effect (LME) based statistical analysis is performed to see the language discriminative capability of the computed features.

\begin{figure}[h!]\centering
\begin{center}
\includegraphics[scale=0.7]{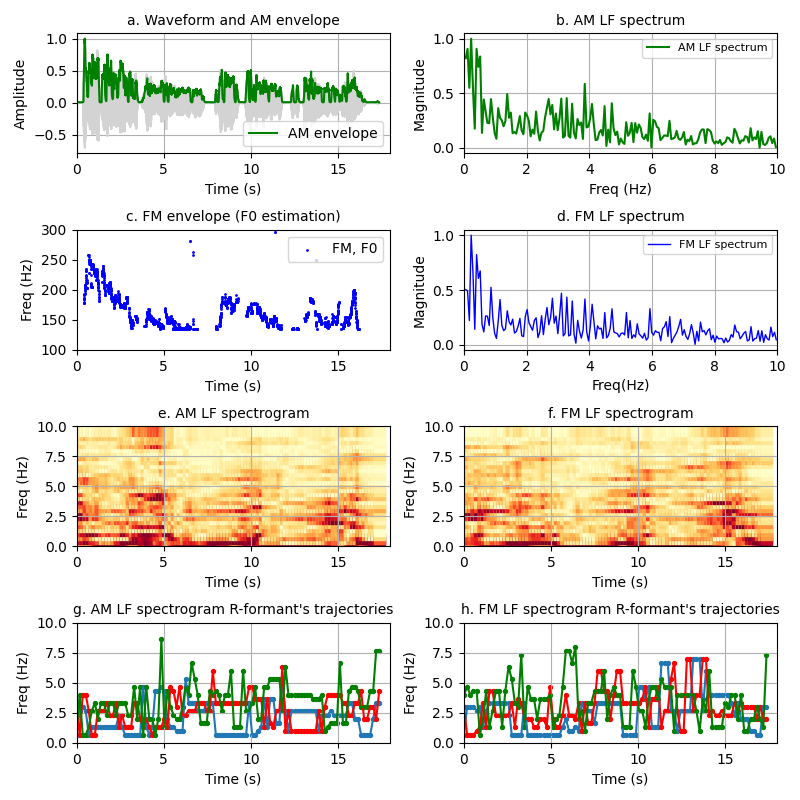}
\caption{Illustration of LF spectrum, LF spectrogram, and temporal trajectories for three dominant R-formants for AM and FM envelopes. a) Speech waveform and AM envelope, b) AM LF spectrum, c) F0 contour, d) FM LF spectrum, e) AM LF spectrogram, f) FM LF spectrogram, g) three dominant R-formant trajectories computed from AM LF spectrogram, and h) three dominant R-formant trajectories computed from AM LF spectrogram.}\label{fig:fmammain}
\end{center}
\end{figure}

In the present experiment, the AM envelope is computed using the Hilbert transform technique. We first normalize the signal using its maximum absolute value and then apply the absolute Hilbert transform to obtain the AM envelope. After smoothing the envelope, a moving window of duration $\it{n}$ sec is used to derive the fast Fourier transform (FFT) of the AM envelope. Here, we vary the $\it{n}$ value from 3 sec to 5 sec with 50 overlapping steps to capture smaller gradual rhythm changes~\cite{gibbon2021jipa}. Gibbon uses a 3-sec duration to compute the LF spectrogram~\cite{gibbon2021jipa}. The resulting AM LF spectrum, computed from each $\it{n}$-second chunk of AM envelopes, is stacked to form a time-frequency representation of long-term rhythm variation, also known as the AM LF spectrograms (Figure~\ref{fig:fmammain} (e)). The LF spectrograms are computed within the 0 to 10 Hz frequency range. We discard the 0 Hz component (DC part) and normalize the spectrum's amplitude. Spectral frequencies having the six highest magnitude peak values are extracted for each LF spectrum (i.e. R-formants)~\cite{gibbon2020storyreading} \cite{gibbonComp}, and the R-formant trajectories are composed along the time axis. The peak picking algorithm is used in the work~\cite{2020SciPy-NMeth-original}. LF spectrograms are therefore used to record changes in frequency distribution over time. In the heatmap pattern of the spectrogram (Figure~\ref{fig:fmammain} (e)), we can observe specific patterns of rhythm variations.  On the other hand, the fundamental frequency (F0) contour of the speech signal is extracted followed by a smoothing of the F0 contour to obtain the FM envelope.  While numerous algorithms exist to estimate F0, in this study, we utilize the RAPT pitch tracking algorithm~\cite{talkin1995robust}. This algorithm's implementation is drawn from the $\textit{pysptk}$~\cite{yamamoto2019r9y9} Python package. It is noteworthy that breaks in the FM envelope, which can result from voiceless consonants and pauses, are not overlooked. Rather, they are considered spectrally relevant segments of the signal. To facilitate a more accurate spectral analysis, we adjust the FM envelope to align with the median F0~\cite{gibbon2021jipa}. However, while plotting the figure (Figure~\ref{fig:fmammain} (c)), discontinuities due to voiceless consonants and pauses are normalized to zero in the FM envelope. We have only shown 0.5 sec to 5 sec duration of the FM envelope for better visualization. We apply a similar moving window technique to the FM envelope, as discussed for the AM envelope, to derive the FM LF spectrograms. Figure~\ref{fig:fmammain} (f) shows the FM spectrogram of an utterance from the Pagro group.

\begin{figure}[]
\begin{center}
\includegraphics[scale=0.8]{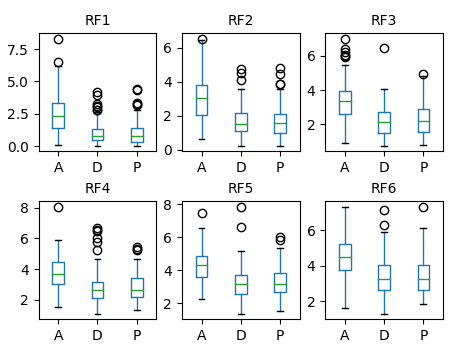}
\caption{Boxplots showing the difference between two dialects of Mising and Assamese for rhythm variance features computed from the R-formant temporal trajectories. Here, A, P, and D
represent Assamese, Pagro, and Delu, respectively, for the AM envelope-based LF spectrogram.}\label{fig:amvarbox}
\end{center}
\end{figure}

\begin{figure}[t]
\begin{center}
\includegraphics[scale=0.8]{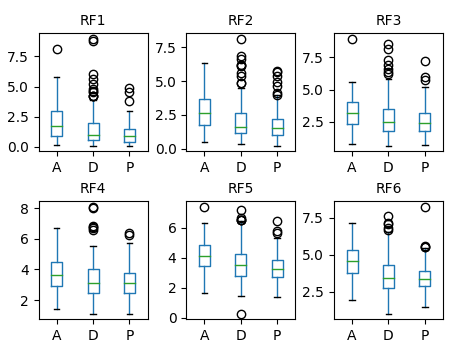}
\caption{Boxplots showing the difference between two dialects of Mising and Assamese for rhythm variance features computed from the R-formant temporal trajectories. Here, A, P, and D
represent Assamese, Pagro, and Delu, respectively, for the FM envelope-based LF spectrogram.}\label{fig:fmvarbox}
\end{center}
\end{figure}

\begin{figure}[t]
\begin{center}
\includegraphics[scale=0.7]{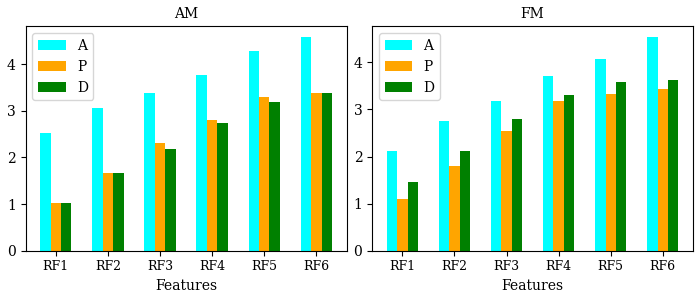}
\caption{The bar plot of variance values of R-formant's temporal trajectories for AM LF spectrogram (left) and FM LF spectrogram (right).}\label{fig:fm-am-bar-var}
\end{center}
\end{figure}

We first study the variance of the R-formant trajectory~\cite{gibbon2022sp} to observe rhythmic variations between Assamese and Mising for both AM and FM envelopes. Figure~\ref{fig:fmammain} (g) and (h) show the three dominant R-formant trajectories for visualization computed from AM and FM LF spectrograms, respectively. For our experiments, we have considered the first six rhythm formants. We calculate the variance of each R-formant trajectory, resulting in six variance-based rhythm values for each utterance. We have plotted the boxplots of six variance-based rhythm measures for Assamese and two varieties of Mising for a 3-second window used to compute the LF spectrogram. Figure~\ref{fig:amvarbox} and Figure~\ref{fig:fmvarbox} show the variance measure's boxplots for AM envelope-based LF spectrogram and FM envelope-based LF spectrogram, respectively. Visual observation of the plots reveals temporal-rhythm variance discrimination between Assamese and Mising dialects, which is more prominent for AM envelope-based measures. The mean value of each R-formant variance measure is computed for both AM and FM LF spectrograms, and the results are shown in Figure~\ref{fig:fm-am-bar-var}. From here, we can see that temporal-rhythm variance is greater for Assamese than Mising, and this difference is greater for Assamese vs. Pagro than for Assamese vs. Delu.
We hypothesized that the variance feature alone may not capture the complete temporal dynamics of the rhythm structure. Therefore, joint spectro-temporal-features-based characterization of the long-term LF spectrogram is explored, and the two-dimensional discrete cosine transform (2D-DCT)-based features are also analyzed. The spectro-temporal variations can be explicitly captured from the time-frequency representation of the rhythm using the 2D-DCT. Earlier, 2D-DCT has been used to extract the joint spectro-temporal features from spectrogram patches for different applications in speech technology~\cite{Bouvrie2008} \cite{SofiaGoodnessMisarticulated2015} \cite{kalita2018intelligibility}. To perform that, the log of the LF spectrogram is computed and then projected to a 2D cosine basis. Let R$_{NM}$ be the LF spectrogram of size $N \times M$, where $N$ and $M$ represent the number of FFT bins and temporal frames (50 in our experiment), respectively. Then, the 2D-DCT $C(k, l)$ of R$_{NM}$ is given as in Equation~\ref{eq1}.
\begin{equation}\label{eq1}
\vspace{-0.5cm}
\frac{2\omega(k)\omega(l)}{\sqrt{NM}}\sum_{a=0}^{M-1}\sum_{b=0}^{N-1}R_{NM}cos\frac{\pi l(2a + 1)}{2M} cos\frac{\pi k(2b + 1)}{2N},
\vspace{0.5cm}
\end{equation}
where, $k = 0, 1, \ldots, N-1$, $l = 0, 1, \ldots, M-1$, 
and 
\[ \omega(k) = \left\{ \begin{array}{ll}
         \frac{1}{\sqrt{2}} & \mbox{if $k = 0$},\\
        1 & \mbox{if $k \neq 0$},\end{array} \right. \] 
        \[ \omega(l) = \left\{ \begin{array}{ll}
         \frac{1}{\sqrt{2}} & \mbox{if $l = 0$},\\
        1 & \mbox{if $l \neq 0$}.\end{array} \right. \]
Then we consider lower-order 2D-DCT coefficients to get a compact representation of the LF spectrogram. In this experiment, we have taken two truncated vertical DCT coefficients and two horizontal DCT coefficients, i.e., $2 \times 2$ dimensional 2D-DCT features. 
$2 \times 2$ matrix is further flattened to make a single vector, termed as 2D-DCT-based features, whose dimension is four. Therefore, in the current study, we have explored six variance-based measures for six dominant R-formant trajectories and four 2D-DCT-based measures, resulting in ten measures. The results of the experiments are confirmed by the LME4 package~\cite{kuznetsova2017lmertest} on R~\cite{team2013r}. 
\section{Results and discussion}\label{results}
Features extracted from the R-formant trajectories and 2D-DCT coefficients are used as feature vectors in the SVM classifiers for classifying Assamese (A) vs. Pagro (P) and Assamese (A) vs. Delu (D). This section describes the results of statistical analysis and SVM-based classifiers.

\begin{table}[h]
\centering
\caption{Pair-wise comparison of the features computed from AM LF spectrogram for Mising and Assamese. $^*$ indicates statistical significance.}\label{table:ampairstats}
\begin{tabular}{|c|c|c|c|c|c|}
\hline
\textbf{Feature selected}   & \textbf{contrast} & \textbf{estimate} & \textbf{SE} & \textbf{df} & \textbf{t-ratio} \\ \hline
Var-RF1-AM & A - D             & 1.7055            & 0.285       & 28.9        & $5.99^*$         \\ \cline{2-6} 
                            & A - P             & 1.638             & 0.286       & 29.1        & $5.73^*$         \\ \hline
Var-RF2-AM & A - D             & 1.4492            & 0.235       & 28.5        & $6.16^*$         \\ \cline{2-6} 
                            & A - P             & 1.366             & 0.237       & 28.8        & $5.77^*$         \\ \hline
Var-RF3-AM & A - D             & 1.312             & 0.235       & 29.3        & $5.58^*$         \\ \cline{2-6} 
                            & A - P             & 1.129             & 0.236       & 29.5        & $4.77^*$         \\ \hline
Var-RF4-AM & A - D             & 1.13              & 0.17        & 23          & $6.65^*$         \\ \cline{2-6} 
                            & A - P             & 0.964             & 0.171       & 23.1        & $5.63^*$         \\ \hline
Var-RF5-AM & A - D             & 1.234             & 0.203       & 29.4        & $6.09^*$         \\ \cline{2-6} 
                            & A - P             & 1.038             & 0.204       & 29.7        & $5.09^*$         \\ \hline
Var-RF6-AM & A - D             & 1.291             & 0.275       & 30.6        & $4.69^*$         \\ \cline{2-6} 
                            & A - P             & 1.189             & 0.276       & 30.9        & $4.30^*$         \\ \hline
2D-DCT1-AM & A - D             & 405               & 313         & 28          & 1.292            \\ \cline{2-6} 
                            & A - P             & 711               & 314         & 28.1        & 2.264            \\ \hline
2D-DCT2-AM & A - D             & -315.8            & 54.7        & 28.6        & $-5.77^*$        \\ \cline{2-6} 
                            & A - P             & -297              & 54.8        & 28.7        & $-5.41^*$        \\ \hline
2D-DCT3-AM & A - D             & -5.26             & 41.6        & 14.11       & -0.126           \\ \cline{2-6} 
                            & A - P             & -80.99            & 41.9        & 13.61       & -1.932           \\ \hline
2D-DCT4-AM & A - D             & 15.67             & 13.3        & 14.11       & 1.178            \\ \cline{2-6} 
                            & A - P             & 10.64             & 13.4        & 13.61       & 0.794            \\ \hline
\end{tabular}
\end{table}

\begin{table}[t]
\centering
\caption{Pair-wise comparison of the features computed from FM LF spectrogram for Mising and Assamese. $^*$ indicates statistical significance. }\label{table:fmpairstats}
\begin{tabular}{|c|c|c|c|c|c|}
\hline
\textbf{Feature selected}   & \textbf{contrast} & \textbf{estimate} & \textbf{SE} & \textbf{df} & \textbf{t-ratio} \\ \hline
Var-RF1-FM & A - D             & 0.809             & 0.353       & 23.3        & 2.288            \\ \cline{2-6} 
                            & A - P             & 1.095             & 0.356       & 23.8        & $3.074^*$        \\ \hline
Var-RF2-FM & A - D             & 0.822             & 0.365       & 23.4        & 2.25             \\ \cline{2-6} 
                            & A - P             & 1.033             & 0.368       & 23.9        & 2.808            \\ \hline
Var-RF3-FM & A - D             & 0.518             & 0.306       & 22.6        & 1.692            \\ \cline{2-6} 
                            & A - P             & 0.734             & 0.31        & 23.3        & 2.37             \\ \hline
Var-RF4-FM & A - D             & 0.467             & 0.186       & 21.8        & 2.511            \\ \cline{2-6} 
                            & A - P             & 0.605             & 0.19        & 23          & $3.182^*$        \\ \hline
Var-RF5-FM & A - D             & 0.513             & 0.164       & 20.9        & $3.132^*$        \\ \cline{2-6} 
                            & A - P             & 0.769             & 0.168       & 22.1        & $4.575^*$        \\ \hline
Var-RF6-FM & A - D             & 0.914             & 0.248       & 22.9        & $3.689^*$        \\ \cline{2-6} 
                            & A - P             & 1.05              & 0.251       & 23.7        & $4.179^*$        \\ \hline
2D-DCT1-FM & A - D             & 651               & 230         & 23.9        & 2.825            \\ \cline{2-6} 
                            & A - P             & 333               & 231         & 24.2        & 1.437            \\ \hline
2D-DCT2-FM & A - D             & -157.2            & 59.1        & 23.7        & -2.659           \\ \cline{2-6} 
                            & A - P             & -214.2            & 59.4        & 24.1        & $-3.606^*$       \\ \hline
2D-DCT3-FM & A - D             & -20.972           & 29          & 19.4        & -0.724           \\ \cline{2-6} 
                            & A - P             & -21.68            & 29.9        & 20.9        & -0.726           \\ \hline
2D-DCT4-FM & A - D             & -5.351            & 11.5        & 19.4        & -0.466           \\ \cline{2-6} 
                            & A - P             & -4.662            & 11.9        & 20.9        & -0.393           \\ \hline
\end{tabular}
\end{table}

\begin{table}[t]
\centering
\caption{Mean accuracy and standard deviation of 3-fold cross-validation for Assamese (A) vs. Pagro (P) and Assamese (A) vs. Delu (D) in cases of AM envelope-based 2DDCT-based features for different durations of window size to compute the AM LF spectrogramm. Classifier: SVM}\label{am-duration-svm}
\resizebox{\textwidth}{!}{%
\begin{tabular}{|c|ccc|ccc|}
\hline
Window duration & \multicolumn{3}{c|}{A vs. P}                                                                     & \multicolumn{3}{c|}{A vs. D}                                                                      \\ \cline{2-7} 
                                 & \multicolumn{1}{c|}{Accuracy (\%)}    & \multicolumn{1}{c|}{F1 score (\%) A}  & F1 score (\%) P  & \multicolumn{1}{c|}{Accuracy (\%)}    & \multicolumn{1}{c|}{F1 score (\%) A}   & F1 score (\%) D  \\ \hline
3 sec                            & \multicolumn{1}{c|}{93.89 $\pm$ 4.72} & \multicolumn{1}{c|}{94.05 $\pm$ 2.49} & 92.50 $\pm$ 7.33 & \multicolumn{1}{c|}{84.17 $\pm$ 3.59} & \multicolumn{1}{c|}{81.71 $\pm$ 7.86}  & 82.98 $\pm$ 7.51 \\ \hline
4 sec                            & \multicolumn{1}{c|}{93.12 $\pm$ 3.88} & \multicolumn{1}{c|}{93.01 $\pm$ 1.77} & 91.88 $\pm$ 6.90 & \multicolumn{1}{c|}{80.48 $\pm$ 2.53} & \multicolumn{1}{c|}{77.96 $\pm$ 8.51}  & 79.53 $\pm$ 4.83 \\ \hline
5 sec                            & \multicolumn{1}{c|}{92.73 $\pm$ 3.70} & \multicolumn{1}{c|}{92.53 $\pm$ 1.71} & 91.56 $\pm$ 6.70 & \multicolumn{1}{c|}{77.89 $\pm$ 4.31} & \multicolumn{1}{c|}{75.85 $\pm$ 10.10} & 76.82 $\pm$ 4.40 \\ \hline
\end{tabular}}
\end{table}

\subsection{Statistical analysis}
All the computed measures for AM and FM envelopes undergo statistical analysis using the Linear Mixed Effects (LME) model. This experiment analyzes six variance-based measures from six R-formants (Var-RFi-AM and Var-RFi-FM, where i = 1 to 6) and four 2D-DCT coefficients. For LME, we consider the computed measures, such as Var-RF1-AM, as the dependent variable. The two fixed factors are gender and language. The LME model considers speaker information as a random factor. To see the variance of factors in the LME models, the type II Wald $\chi^2$ tests using the car package~\cite{fox2018r} are conducted. The posthoc Bonferroni tests are carried out using the Emmeans package~\cite{lenth2018emmeans} to examine the pair-wise differences of the categories. The p-value of Anova results for all six variance measures is found to be a significant p-value at $<0.01$ for the AM LF spectrogram. Among the four computed 2D-DCT-based features, a significant p-value is found for the second one (2D-DCT2) for both AM and FM LF spectrograms. Other coefficients of 2D-DCT are found to be insignificant in the post-hoc tests. Table~\ref{table:ampairstats} shows the results of the pair-wise test only for the cases having significant p-values for AM-based measures. Table~\ref{table:fmpairstats} tabulates the pair-wise comparison of the features computed in the FM LF spectrogram. The table shows that the variance values computed from the first, fourth, fifth, and sixth R-formant trajectories have a statistically significant p-value at $<0.01$ for Assamese vs. Pagro. Similar to the AM LF spectrogram-based 2DDCT measures, only the second 2DDCT coefficient has a statistically significant p-value at $<0.01$ for the FM LF spectrogram.

\subsection{SVM-based classification and results}
We develop SVM-based classification systems for Pagro (P) vs. Assamese (P) and Delu (D) vs. Assamese (A) to determine if the derived features effectively discriminate language. Here, training is done in a speaker-independent manner with 3-fold cross-validation. Each iteration uses $80\%$ of total data for training and $20\%$ for testing. We then calculate the average accuracy and F1 score on the test sets for each fold as evaluation metrics. The grid-search method has been applied to tune the SVM hyperparameters, such as C and $\gamma$. Table~\ref{svm_1} summarizes the classification results of the 3-fold cross-validation. F1-A and F1-P (F1-D) represent the positive classes in the F1-score for Assamese and Pagro (Delu), respectively.

\begin{table}[]
\centering
\caption{Mean accuracy and standard deviation of 3-fold cross-validation for Assamese (A) vs. Pagro (P) and Assamese (A) vs. Delu (D) in cases of FM envelope-based 2DDCT-based features for different durations of window size to compute the FM LF spectrogramm. Classifier: SVM}\label{fm-duration-svm}
\resizebox{\textwidth}{!}{%
\begin{tabular}{|c|ccc|ccc|}
\hline
Window duration & \multicolumn{3}{c|}{A vs. P}                                                                       & \multicolumn{3}{c|}{A vs. D}                                                                        \\ \cline{2-7} 
                                 & \multicolumn{1}{c|}{Accuracy (\%)}     & \multicolumn{1}{c|}{F1 score (\%) A}   & F1 score (\%) P  & \multicolumn{1}{c|}{Accuracy (\%)}     & \multicolumn{1}{c|}{F1 score (\%) A}   & F1 score (\%) D   \\ \hline
3 sec                            & \multicolumn{1}{c|}{89.63 $\pm$ 5.3}   & \multicolumn{1}{c|}{89.33 $\pm$ 4.41}  & 87.93 $\pm$ 9.14 & \multicolumn{1}{c|}{79.25 $\pm$ 8.97}  & \multicolumn{1}{c|}{78.15 $\pm$ 12.46} & 78.64 $\pm$ 5.86  \\ \hline
4 sec                            & \multicolumn{1}{c|}{86.24 $\pm$ 5.24}  & \multicolumn{1}{c|}{85.29 $\pm$ 4.00}  & 85.33 $\pm$ 8.85 & \multicolumn{1}{c|}{75.34 $\pm$ 6.37}  & \multicolumn{1}{c|}{73.07 $\pm$ 10.94} & 75.35 $\pm$ 6.85  \\ \hline
5 sec                            & \multicolumn{1}{c|}{67.67 $\pm$ 10.53} & \multicolumn{1}{c|}{56.99 $\pm$ 21.24} & 72.94 $\pm$ 7.14 & \multicolumn{1}{c|}{56.54 $\pm$ 24.87} & \multicolumn{1}{c|}{60.88 $\pm$ 18.19} & 49.59 $\pm$ 35.23 \\ \hline
\end{tabular}}
\end{table}

\begin{table*}[]
\centering
\caption{Mean accuracy and standard deviation of 3-fold cross-validation for Assamese (A) vs. Pagro (P) and Assamese (A) vs. Delu (D) in case of R-Formants variance and 2DDCT-based features. Classifier: SVM}
\resizebox{\textwidth}{!}{%
\begin{tabular}{|c|ccc|ccc|}
\hline
Feature (Dimension) & \multicolumn{3}{c|}{A vs. P}                                                                       & \multicolumn{3}{c|}{A vs. D}                                                                       \\ \cline{2-7} 
                                     & \multicolumn{1}{c|}{Accuracy (\%)}    & \multicolumn{1}{c|}{F1 score (\%) (A)} & F1 score (\%) (P) & \multicolumn{1}{c|}{Accuracy (\%)}    & \multicolumn{1}{c|}{F1 score (\%) (A)} & F1 score (\%) (D) \\ \hline
MFCCs (78)                     & \multicolumn{1}{c|}{92.32 $\pm$ 6.57} & \multicolumn{1}{c|}{91.79 $\pm$ 6.09} & 91.56 $\pm$ 5.20                           & \multicolumn{1}{c|}{88.79 $\pm$ 5.70} & \multicolumn{1}{c|}{88.19 $\pm$ 2.0}   & \multicolumn{1}{c|}{88.45 $\pm$ 2.5}       \\ \hline
VarAMRF1 (1)                           & \multicolumn{1}{c|}{77.06 $\pm$ 7.80} & \multicolumn{1}{c|}{71.37 $\pm$ 6.10}  & 79.71 $\pm$ 9.94  & \multicolumn{1}{c|}{73.79 $\pm$ 9.80} & \multicolumn{1}{c|}{67.39 $\pm$ 4.60}  & 76.01 $\pm$ 1.36  \\ \hline
VarAMRF1-VarAMRF6 (6)                          & \multicolumn{1}{c|}{79.48 $\pm$ 3.07} & \multicolumn{1}{c|}{76.57 $\pm$ 2.85}  & 80.51 $\pm$ 6.11  & \multicolumn{1}{c|}{76.70 $\pm$ 3.37} & \multicolumn{1}{c|}{72.52 $\pm$ 6.03}  & 77.38 $\pm$ 8.43  \\ \hline
VarFMRF1 (1)                           & \multicolumn{1}{c|}{68.45 $\pm$ 4.93} & \multicolumn{1}{c|}{65.45 $\pm$ 3.53}  & 68.90 $\pm$ 3.73  & \multicolumn{1}{c|}{66.45 $\pm$ 5.44} & \multicolumn{1}{c|}{64.45 $\pm$ 2.44}  &   67.33 $\pm$ 3.67\\ \hline
VarFMRF1-VarFMRF6 (6)                          & \multicolumn{1}{c|}{71.23 $\pm$ 3.21} & \multicolumn{1}{c|}{69.11 $\pm$ 3.56}  & 72.11 $\pm$ 4.14  & \multicolumn{1}{c|}{68.45 $\pm$ 3.77} & \multicolumn{1}{c|}{67.22 $\pm$ 4.67}  &   70.88 $\pm$ 4.23\\ \hline
2D-DCT-AM (4)                            & \multicolumn{1}{c|}{93.89 $\pm$ 4.72} & \multicolumn{1}{c|}{94.05 $\pm$ 2.49}  & 92.50 $\pm$ 7.33  & \multicolumn{1}{c|}{84.17 $\pm$ 3.59} & \multicolumn{1}{c|}{81.71 $\pm$ 7.86}  & 82.98 $\pm$ 7.51  \\ \hline
2D-DCT-FM (4)                            & \multicolumn{1}{c|}{86.24 $\pm$ 5.24} & \multicolumn{1}{c|}{85.29 $\pm$ 4.00}  & 85.33 $\pm$ 8.85 & \multicolumn{1}{c|}{ 79.25 $\pm$ 8.97} & \multicolumn{1}{c|}{78.15 $\pm$ 12.46}  & 78.64 $\pm$ 5.86  \\ \hline
2D-DCT-AM + 2D-DCT-FM (8)                            & \multicolumn{1}{c|}{94.30 $\pm$ 3.86} & \multicolumn{1}{c|}{94.34 $\pm$ 2.24}  & 93.19 $\pm$ 6.36 & \multicolumn{1}{c|}{ 78.31 $\pm$ 5.93} & \multicolumn{1}{c|}{76.30 $\pm$ 11.41}  & 77.34 $\pm$ 3.21   \\ \hline
\end{tabular}}
\label{svm_1}
\end{table*}
We employ Mel-frequency cepstral coefficients (MFCCs) with SVM models in the initial setup to serve as a baseline language classification system for this work. Here, each speech signal is short-term processed with a 20 msec Hamming window with a 10 msec frameshift, and MFCCs for each frame are computed. An energy-based voice activity detection is performed to obtain the voiced region from the speech, and the MFCC features corresponding to the voiced regions are retained. In this experiment, we have considered the default 13-dimensional MFCCs, and their $\Delta$ and $\Delta\Delta$ are also computed to model the temporal context. Therefore, the total dimension of the MFCC feature vector becomes 39. The mean and standard deviation of each feature are computed for all the frames to model the temporal statistics and to generate one vector for each utterance of dimension 78 (39-dimensional mean and 39-dimensional standard deviation vectors). This vector acts as an input to the SVM model for classification. The results from a 3-fold cross-validation using MFCCs, presented in Table~\ref{svm_1}, indicate that Assamese vs. Pagro has a marginally better performance than Assamese vs. Delu. We achieved an accuracy of 92.32\% for A vs. P and 88.79\% for A vs. D.

We have computed the variance of R-formant trajectories from the LF spectrogram derived using different window durations, such as 3 sec, 4 sec, and 5 sec. However, we find that variance features computed from the LF spectrogram with a 3-second window duration provide better performance for both AM and FM envelopes. The results for variance-based measures are noted in Table~\ref{svm_1}. The Var-RF1-AM variance feature calculated from the AM LF spectrogram gives an accuracy of $77.06\%$ for P vs. A and $73.79\%$ percent for D vs. A (see Table~\ref{svm_1}). However, the standard deviation of accuracy and F1-score are higher, which means that Var-RF1-AM has not performed well consistently for all the folds. All six combined features, from Var-RF1-AM to Var-RF2-AM, show a similar trend across all evaluation metrics. A slight improvement in accuracy ($79.48\%$) and F1 score ($76.57\%$ and $80.51\%$) is observed for the classification P vs. A, with low standard deviation values of accuracy and F1 score. In both systems, the F1 score values for Assamese are lower than those for Pagro and Delu. The performance of FM-based variance measures (Var-FM-RF1 to Var-RF6-FM) is comparatively less than that of the AM counterpart. In both cases, P vs. A performs better than D vs. A; therefore, the temporal variability of R-formants in Assamese is closer to Delu than to Pagro.

The results of different window durations of the envelope for computing AM and FM LF spectrograms are shown in Figure~\ref{am-duration-svm} and Figure~\ref{fm-duration-svm} for 2D-DCT-based measures, respectively. It is evident that a 3-second window duration yields the best results for both AM and FM LF spectrograms, and an increase in window duration leads to a decline in performance. One possible reason is that our speech files have an average duration of 12.94 seconds; therefore, a longer window duration may not properly capture the finer rhythm variations in the LF spectrogram. However, there is scope for analyzing the effect of duration while computing the LF spectrogram for the longer-duration speech files, especially to understand how rhythm is related to phrase and discourse. In the future, we intend to conduct a thorough investigation in this area. While inputting AM envelope-based 2D-DCT features, the model performs significantly better than the variance-based features (Table~\ref{svm_1}). We observe an accuracy of $93.89\%$ for P vs. A and an accuracy of $84.17\%$ for D vs. A in the case of the AM LF spectrogram. However, the FM-based 2D-DCT feature shows less accuracy than the AM-based feature. This means that the AM envelope encodes more discriminable information when classifying Assamese vs. Mising. Finally, we perform the feature-level fusion of both AM- and FM-based 2D-DCT features and result in an improvement in accuracy ($94.30\%$) for A vs. P, while $78.31\%$ for A vs. D. The classification performance of P vs. A is superior to that of D vs. A, which may imply that Assamese is rhythmically closer to Delu than Pagro.

\section{Summary and conclusions}\label{discussion}
We have attempted to investigate the long-term rhythm characteristics encoded in the AM and FM LF spectrograms. We derive these LF spectrograms for an utterance by introducing a moving FFT window on the AM or FM envelope. We have explored the technique on a spontaneous speech dataset of 19 native speakers to contrast Assamese and Mising rhythms using features derived from LF spectrograms. In this work, we explore the variance of rhythm formants and the 2D-DCT-based representation of the LF spectrogram as rhythm corelates. As a case study, the derived features are also used to build the SVM-based classifier for Assamese vs. Pagro and Assamese vs. Delu. An accuracy of $93.89\%$ for Pagro vs. Assamese and $84.17\%$ for Delu vs. Assamese using AM LF spectrogram-based 2D-DCT features is obtained. We combining with the FM LF spectrogram-based features, we get $94.30\%$ for the Pagro vs. Assamese classification. 

Interestingly, the performance of Assamese vs. Pagro is better than Delu vs. Assamese. In another sense, Assamese is rhythmically closer to Pagro than Delu - an observation can also be made from the statistical analysis, where the contrast estimate for Assamese vs. Pagro is more than Assamese vs. Delu for all the features. Since Assamese is the lingua franca of Assam, its influence is one of the significant factors bringing out rhythmic variations in Pagro and Delu. The degree of influence on Mising dialects varies concerning the degree of influence from the Assamese language and different geographical settings. However, it is found that none of these features conveys any discriminative cues to classify Pagro and Delu. A detailed analysis is required with more speech data from different speakers to validate the claimed point, and future work is planned in this direction.  We can model the LF spectrograms using a convolutional neural network instead of a 2D-DCT-based representation, and also understand the rhythm's encoding in the LF spectrogram with more training data.

\section{Acknowledgement}\label{Ack}
The Mising and Assamese speakers from the several Upper Assamese villages who willingly participated in this effort are gratefully acknowledged by the authors.

\bibliographystyle{unsrt}  
\bibliography{mybib}

\begin{thebibliography}{10}

\bibitem{Gibbon2019quantify}
Dafydd Gibbon and Peng Li.
\newblock Quantifying and correlating rhythm formants in speech.
\newblock In {\em Proc. Linguistic Patterns in Spontaneous Speech (LPSS)}. Taipei, Academia Sinica, 2019.

\bibitem{gibbon2021jipa}
Dafydd Gibbon.
\newblock The rhythms of rhythm.
\newblock {\em Journal of the International Phonetic Association}, page 1–33, 2021.

\bibitem{Nespor2019}
Marina Nespor.
\newblock {\em On the rhythm parameter in phonology}, pages 157--176.
\newblock De Gruyter Mouton, 1990.

\bibitem{RAMUS}
Franck Ramus, Marina Nespor, and Jacques Mehler.
\newblock Correlates of linguistic rhythm in the speech signal.
\newblock {\em Cognition}, 75(1):AD3--AD30, 2000.

\bibitem{moraJ}
Robert~F Port, Jonathan Dalby, and Michael O’Dell.
\newblock Evidence for mora timing in {J}apanese.
\newblock {\em The Journal of the Acoustical Society of America}, 81(5):1574--1585, 1987.

\bibitem{murty2007perceptual}
Lalita Murty, Takashi Otake, and Anne Cutler.
\newblock Perceptual tests of rhythmic similarity: I. mora rhythm.
\newblock {\em Language and Speech}, 50(1):77--99, 2007.

\bibitem{dihingia2020rhythm}
Leena Dihingia and Priyankoo Sarmah.
\newblock Rhythm and speaking rate in {A}ssamese varieties.
\newblock In {\em Proc. 10th International Conference on Speech Prosody 2020}, pages 561--565, 2020.

\bibitem{dauer1983stress}
Rebecca~M Dauer.
\newblock Stress-timing and syllable-timing reanalyzed.
\newblock {\em Journal of phonetics}, 11(1):51--62, 1983.

\bibitem{GrabeLow}
Esther Grabe and Ee~Ling Low.
\newblock {\em Durational variability in speech and the Rhythm Class Hypothesis}, pages 515--546.
\newblock De Gruyter Mouton, 2008.

\bibitem{low}
Low~Ee Ling, Esther Grabe, and Francis Nolan.
\newblock Quantitative characterizations of speech rhythm: Syllable-timing in {S}ingapore {E}nglish.
\newblock {\em Language and Speech}, 43(4):377--401, 2000.

\bibitem{dellwo2003}
Volker Dellwo and Petra Wagner.
\newblock Relations between language rhythm and speech rate.
\newblock 07 2015.

\bibitem{ferragne2004rhythm}
Emmanuel Ferragne and Fran{\c{c}}ois Pellegrino.
\newblock Rhythm in read {B}ritish {E}nglish: interdialect variability.
\newblock In {\em Eighth International Conference on Spoken Language Processing}, 2004.

\bibitem{Dellwo2015jasa}
V.~Dellwo, A.~Leemann, and M.~J. Kolly.
\newblock Rhythmic variability between speakers: Articulatory, prosodic, and linguistic factors.
\newblock {\em The Journal of the Acoustical Society of America}, 137:1513–1528, 2015.

\bibitem{praat}
Paul Boersma and David Weenink.
\newblock Praat: doing phonetics by computer (version 5.1.13), 2009.

\bibitem{wiget}
Lukas Wiget, Laurence White, Barbara Schuppler, Izabelle Grenon, Olesya Rauch, and Sven~L Mattys.
\newblock How stable are acoustic metrics of contrastive speech rhythm?
\newblock {\em The Journal of the Acoustical Society of America}, 127(3):1559--1569, 2010.

\bibitem{Loukina}
Anastassia Loukina, Greg Kochanski, Burton Rosner, Elinor Keane, and Chilin Shih.
\newblock Rhythm measures and dimensions of durational variation in speech.
\newblock {\em The Journal of the Acoustical Society of America}, 129(5):3258--3270, 2011.

\bibitem{Todd1994}
Neil P.~McAngus Todd.
\newblock The auditory “primal sketch”: A multiscale model of rhythmic grouping.
\newblock {\em Journal of New Music Research}, 23(1):25--70, 1994.

\bibitem{cummins1999language}
Fred Cummins, Felix Gers, and J{\"u}rgen Schmidhuber.
\newblock Language identification from prosody without explicit features.
\newblock In {\em Sixth European Conference on Speech Communication and Technology}, 1999.

\bibitem{traunmuller1994conventional}
Hartmut Traunm{\"u}ller.
\newblock Conventional, biological and environmental factors in speech communication: A modulation theory.
\newblock {\em Phonetica}, 51(1-3):170--183, 1994.

\bibitem{Kohler2009}
K.~J. Kohler.
\newblock Whither speech rhythm research?
\newblock {\em Phonetica}, 66:5--14, 2009.

\bibitem{Arvanitiusefulness2012}
Amalia Arvaniti.
\newblock The usefulness of metrics in the quantification of speech rhythm.
\newblock {\em J. Phonetics}, 40:351, 2012.

\bibitem{CUMMINS1998145}
Fred Cummins and Robert Port.
\newblock Rhythmic constraints on stress timing in english.
\newblock {\em Journal of Phonetics}, 26(2):145--171, 1998.

\bibitem{Tilsen2013}
S.~Tilsen.
\newblock Speech rhythm analysis with decomposition of the amplitude envelope: Characterizing rhythmic patterns within and across languages.
\newblock {\em J. Acoust. Soc. Am.}, 134:628, 2013.

\bibitem{nature}
David Poeppel and M.~Florencia Assaneo.
\newblock Speech rhythms and their neural foundations.
\newblock {\em Nature Reviews Neuroscience}, 21(6):322--334, 2020.

\bibitem{ding2017temporal}
Nai Ding, Aniruddh~D Patel, Lin Chen, Henry Butler, Cheng Luo, and David Poeppel.
\newblock Temporal modulations in speech and music.
\newblock {\em Neuroscience \& Biobehavioral Reviews}, 81:181--187, 2017.

\bibitem{barbosa2002explaining}
Pl{\'\i}nio~A Barbosa.
\newblock Explaining cross-linguistic rhythmic variability via a coupled-oscillator model of rhythm production.
\newblock In {\em Speech Prosody 2002, International Conference}, 2002.

\bibitem{inden2012rapid}
Benjamin Inden, Zofia Malisz, Petra Wagner, and Ipke Wachsmuth.
\newblock Rapid entrainment to spontaneous speech: A comparison of oscillator models.
\newblock In {\em Proceedings of the Annual Meeting of the Cognitive Science Society}, volume~34, 2012.

\bibitem{O'Dell}
Michael O'Dell and Tommi Nieminen.
\newblock Coupled oscillator model of speech rhythm.
\newblock In J.~Ohala, Y.~Hasegawa, M.~Ohala, D.~Granville, and A.~Bailey, editors, {\em Proceedings of The XIVth International Congress of Phonetic Sciences}, volume Vol. 2, pages 1075--1078. University of California, 1999.

\bibitem{Tilsen2008LowfrequencyFA}
Sam Tilsen and Keith Johnson.
\newblock Low-frequency fourier analysis of speech rhythm.
\newblock {\em The Journal of the Acoustical Society of America}, 124 2:EL34--9, 2008.

\bibitem{gibbonComp}
Dafydd Gibbon.
\newblock Computational induction of prosodic structure.
\newblock {\em Studies in Prosodic Grammar}, 6 (2):1--46, 2020.

\bibitem{gibbon2018future}
Dafydd Gibbon.
\newblock The future of prosody: It's about time.
\newblock In {\em Proc. 9th International Conference on Speech Prosody 2018}, pages 1--9, 2018.

\bibitem{gibbon2020storyreading}
Dafydd Gibbon.
\newblock Rhythm formants of story reading in standard {M}andarin.
\newblock {\em {C}hinese Journal of Phonetics}, 14:1--16, 2020.

\bibitem{kaustubh2023rhythm}
Kumar Kaustubh, Parismita Gogoi, and SRM Prasanna.
\newblock Rhythm formant analysis for automatic depression classification.
\newblock In {\em International Conference on Speech and Computer}, pages 94--106. Springer, 2023.

\bibitem{taid}
T.~Taid.
\newblock A short note on {M}ising phonology.
\newblock {\em Linguistics of the Tibeto-Burman Area}, 10.1, 1987.

\bibitem{Prasad}
B.R. Prasad.
\newblock {M}ising grammar.
\newblock {\em Mysore, Central Institute of Indian languages (CIIL)}, Eds: Sastry and Abraham, 1991.

\bibitem{Goswami-and-Tamuli-2003}
G.~C. Goswami and Jyotiprakash Tamuli.
\newblock Asamiya.
\newblock In George Cardona and Dhanesh Jain, editors, {\em The {I}ndo-{A}ryan Languages}, pages 391--443. Routledge, London, 2003.

\bibitem{Rajeev}
Rajeev~K. Doley.
\newblock Can acculturation lead to language death? a case study.
\newblock {\em IOSR Journal Of Humanities And Social Science}, 9:33--36, 2013.

\bibitem{Loreina}
Loreina Pagag.
\newblock Phonological processes in {M}ising language: A privilege theoretic account.
\newblock In {\em Proc. The European Conference on Language Learning}, pages 1--9, 2015.

\bibitem{parismita_specom}
Parismita Gogoi, Priyankoo Sarmah, and S.~R.~M. Prasanna.
\newblock Automatic rhythm and speech rate analysis of {M}ising spontaneous speech.
\newblock In S.~R.~Mahadeva Prasanna, Alexey Karpov, K.~Samudravijaya, and Shyam~S. Agrawal, editors, {\em Speech and Computer}, pages 201--213, Cham, 2022. Springer International Publishing.

\bibitem{parismitatallip}
Parismita Gogoi, Priyankoo Sarmah, and S.~R.~M. Prasanna.
\newblock Cross-linguistic rhythm analysis of {M}ising and {A}ssamese.
\newblock {\em ACM Trans. Asian Low-Resour. Lang. Inf. Process.}, 2024.
\newblock Just Accepted.

\bibitem{gibbonRhythmZoneTheory}
Dafydd Gibbon and Xuewei Lin.
\newblock Rhythm zone theory: Speech rhythms are physical after all.
\newblock {\em Approaches to the Study of Sound Structure and Speech. Interdisciplinary Work in Honour of Katarzyna Dziubalska-Kołaczyk}, 2020.

\bibitem{2020SciPy-NMeth-original}
Pauli Virtanen, Ralf Gommers, Travis~E. Oliphant, Matt Haberland, Tyler Reddy, David Cournapeau, Evgeni Burovski, Pearu Peterson, Warren Weckesser, Jonathan Bright, St{\'e}fan~J. {van der Walt}, Matthew Brett, Joshua Wilson, K.~Jarrod Millman, Nikolay Mayorov, Andrew R.~J. Nelson, Eric Jones, Robert Kern, Eric Larson, C~J Carey, {\.I}lhan Polat, Yu~Feng, Eric~W. Moore, Jake {VanderPlas}, Denis Laxalde, Josef Perktold, Robert Cimrman, Ian Henriksen, E.~A. Quintero, Charles~R. Harris, Anne~M. Archibald, Ant{\^o}nio~H. Ribeiro, Fabian Pedregosa, Paul {van Mulbregt}, and {SciPy 1.0 Contributors}.
\newblock {{SciPy} 1.0: Fundamental Algorithms for Scientific Computing in Python}.
\newblock {\em Nature Methods}, 17:261--272, 2020.

\bibitem{talkin1995robust}
David Talkin and W~Bastiaan Kleijn.
\newblock A robust algorithm for pitch tracking (rapt).
\newblock {\em Speech coding and synthesis}, 495:518, 1995.

\bibitem{yamamoto2019r9y9}
Ryuichi Yamamoto, Joao Felipe, and Merlijn Blaauw.
\newblock r9y9/pysptk: 0.1. 14.
\newblock {\em URL: https://github. com/r9y9/pysptk}, 2019.

\bibitem{gibbon2022sp}
Dafydd Gibbon.
\newblock Speech rhythms: learning to discriminate speech styles.
\newblock In {\em Proc. International Conference on Speech Prosody 2022}, pages 302--306, 05 2022.

\bibitem{Bouvrie2008}
J.~Bouvrie, T.~Ezzat, and T.~Poggio.
\newblock Localized spectro-temporal cepstral analysis of speech.
\newblock In {\em 2008 IEEE International Conference on Acoustics, Speech and Signal Processing}, pages 4733--4736, March 2008.

\bibitem{SofiaGoodnessMisarticulated2015}
Sofia Str\"{o}mbergsson, Giampiero Salvi, and David House.
\newblock Acoustic and perceptual evaluation of category goodness of /t/ and /k/ in typical and misarticulated children's speech.
\newblock {\em The Journal of the Acoustical Society of America}, 137(6):3422--3435, 2015.

\bibitem{kalita2018intelligibility}
Sishir Kalita, SR~Mahadeva~Prasanna, and S~Dandapat.
\newblock Intelligibility assessment of cleft lip and palate speech using gaussian posteriograms based on joint spectro-temporal features.
\newblock {\em The Journal of the Acoustical Society of America}, 144(4):2413--2423, 2018.

\bibitem{kuznetsova2017lmertest}
Alexandra Kuznetsova, Per~B Brockhoff, and Rune~HB Christensen.
\newblock lmertest package: tests in linear mixed effects models.
\newblock {\em Journal of statistical software}, 82:1--26, 2017.

\bibitem{team2013r}
R~Core Team et~al.
\newblock R: A language and environment for statistical computing.
\newblock 2013.

\bibitem{fox2018r}
John Fox and Sanford Weisberg.
\newblock {\em An R companion to applied regression}.
\newblock Sage publications, 2018.

\bibitem{lenth2018emmeans}
Russell Lenth, Henrik Singmann, Jonathon Love, Paul Buerkner, and Maxime Herve.
\newblock Emmeans: Estimated marginal means, aka least-squares means.
\newblock {\em R package version}, 1(1):3, 2018.

\end{thebibliography}

\end{document}